\documentclass{ws-jai}
\usepackage[flushleft]{threeparttable}
\usepackage{graphicx}
\usepackage{caption}
\usepackage{subcaption}
\usepackage{multirow}
\usepackage{tikz}
\usepackage{xfrac}
\usetikzlibrary{shapes.geometric, arrows}
\tikzstyle{startstop} = [rectangle, rounded corners, minimum width=3cm, minimum height=1cm,text centered, draw=black, fill=red!30]
\tikzstyle{io} = [trapezium, trapezium left angle=70, trapezium right angle=110, minimum width=3cm, minimum height=1cm, text centered, draw=black, fill=blue!30]
\tikzstyle{decision} = [diamond, minimum width=3cm, minimum height=1cm, text centered, text width=2.0cm, draw=black, fill=green!30]
\tikzstyle{process} = [rectangle, minimum width=3cm, minimum height=1cm, text centered, text width=6.5cm, draw=black, fill=orange!30]
\tikzstyle{arrow} = [thick,->,>=stealth]

\begin{document}

\catchline{}{}{}{}{} 

\markboth{O. Restrepo et al.}{Optimization of Antenna Performance for Global 21-cm Observations and Verification Using Scaled Copies
}

\title{Optimization of Antenna Performance for Global 21-cm Observations and Verification Using Scaled Copies
}

\author{O. A. Restrepo$^{1,2}$, F. I. Lucero$^{1}$, G. Chaparro$^{3}$, R. Rodriguez$^{4}$, F. Pizarro $^{5}$, R. Bustos $^{6}$, M. D\'iaz $^{6}$, and F. P. Mena$^{1,7}$}

\address{
$^{1}$Departamento de Ingeniería Eléctrica, Universidad de Chile, Santiago, Chile, orestrepo@ug.uchile.cl\\
$^{2}$Universidad ECCI, Bogot\'a, Colombia, orestrepog@ecci.edu.co\\
$^{3}$Universidad de Antioquia, Medell\'in, Colombia\\
$^{4}$Institute of Electricity and Electronics, Universidad Austral de Chile, General Lagos 2086, Campus Miraflores, Valdivia, Región de los Ríos, Chile\\
$^{5}$Escuela de Ingenier\'ia El\'ectrica, Pontificia Universidad Cat\'olica de Valpara\'iso\\
$^{6}$Departamento de Ingenier\'ia El\'ectrica, Universidad Cat\'olica de la Sant\'isima Concepci\'on, Alonso de Ribera 2850, Concepci\'on, Chile \\
$^{7}$Central Development Laboratory, National Radio Astronomy Observatory, Charlottesville, VA, USA 
}

\maketitle

\corres{$^{1}$ orestrepo@ug.uchile.cl}

\begin{history}
\received{(to be inserted by publisher)};
\revised{(to be inserted by publisher)};
\accepted{(to be inserted by publisher)};
\end{history}

\begin{abstract}

The sky-averaged cosmological 21-cm signal can improve our understanding of the evolution of the early Universe from the Dark Age to the end of the Epoch of Reionization.
Although the EDGES experiment reported an absorption profile of this signal, there have been concerns about the plausibility of these results, motivating independent 
validation experiments. One of these initiatives is the Mapper of the IGM Spin Temperature (MIST), which is planned to be deployed at different remote locations around the 
world. One of its key features is that it seeks to comprehensively compensate for systematic uncertainties through detailed modeling and characterization of its 
different instrumental subsystems, particularly its antenna. Here we propose a novel optimizing scheme which can be used to design an antenna applied to
MIST, improving bandwidth, return loss, and beam chromaticity. This new procedure combines the Particle Swarm Optimization (PSO) algorithm with a commercial electromagnetic simulation 
software (HFSS). We improved the performance of two antenna models: a rectangular blade antenna, similar to the one used in the EDGES experiment, and a trapezoidal bow-tie 
antenna. Although the performance of both antennas improved after applying our optimization method, we found that our bow-tie model outperforms the blade antenna by achieving 
lower reflection losses and beam chromaticity in the entire band of interest. To further validate the optimization process, we also built and characterized 1:20 scale models 
of both antenna types showing an excellent agreement with our simulations.
\end{abstract}


\section{Introduction}

\noindent In our current understanding of the formation of the Universe, the tiny density fluctuations that we now observe as temperature variations in the Cosmic Microwave 
Background (CMB) grew through gravitational instability until they formed the cosmic structures that we see around us today \citep{Zaldarriaga_2004}. 
Once the first stars formed in the transition between these two epochs, known as the Dark Ages \cite{Haemmerl}, their ultraviolet light eventually penetrated the primordial hydrogen gas, altering 
the excitation state of its 21-cm hyperfine line. This alteration would cause the gas to absorb photons from the CMB producing an all-sky distortion of the spectrum that 
would be redshifted to the VHF part of the radio spectrum in our local frame, therefore being observable today \cite{Furlanetto_2006}. 
Several experiments aim to detect the spectral signature of the cosmological redshifted 21-cm line after averaging over the entire sky. For example, \citep[BIGHORNS,][]{Sokolowski2015BIGHORNSB}, \citep[SARAS 2,][]{Singh2017SARAS2A}, \citep[SCI-HI,][]{Voytek_2014}, \citep[LEDA,][]{Bernardi}, \citep[PRI\textsuperscript{Z}M,][]{Philip2018ProbingRI}, \citep[CTP,][]{Nhan}, \citep[REACH,][]{8879199}, \citep[High-Z,][]{Navros}, \citep[SITARA,][]{10.1093/mnras/staa2804}, \citep[ASSASSIN,][]{thekkeppattu_mckinley_trott_jones_ung_2022}, \citep[EDGES,][]{Rogers}. \\

\noindent In 2018, EDGES reported the detection of a 530~mK absorption feature centered at 78.1~MHz \cite{Bowman_2018}. The amplitude of this absorption feature is 2$-$3 times higher than what is expected from current models \cite{Pritchard_2012}; \cite{Fialkov_2014}; \cite{Fialkov_2016}; \cite{10.1093/mnras/stx2065}. 
The reported profile also shows a saturated, optically thick line profile rather than the Gaussian-like profile expected from models \cite{Barkana_2018}.  Moreover, in 2022 the SARAS 3 experiment reported measurements showing that the profile found by EDGES is not of astrophysical origin \cite{singh2022detection}. It has been suggested that the purported signal could be an artifact due to an unmodeled periodic instrumental feature, thus refuting the cosmological origin of the EDGES signal \cite{Hills_2018}, \cite{Castel_2022}. Therefore, this detection needs to be confirmed and validated by independent experiments at different sites in the world. In this context, a new instrument, the Mapper of the IGM Spin Temperature (MIST)\footnote{http://www.physics.mcgill.ca/mist/} is currently being designed and deployed to observe the cosmological 21-cm hydrogen line from several sites such as the Canadian Arctic, Uapishka in Canada, California, Nevada (US), northern Chile, and Patagonia, covering a wide range of latitudes with a single dipole antenna. MIST will observe the sky-averaged brightness temperature in the 40-120 MHz range. One of the goals of MIST is to place particular attention to the expected frequency performance of the deployed antenna in order to ensure that its chromaticity does not introduce spurious effects that may confuse the expected signal. \\ 

\noindent It is crucial to have a comprehensive understanding of the antenna performance in 21-cm cosmic signal 
experiments due to the expected signal being much weaker than the sky background radiation. To ensure accurate data analysis, the antenna design must be meticulously modeled 
and carefully planned prior to the experiment. The chromaticity of the antenna is a critical aspect in its design, as it can impact the accuracy of measurements and frequency 
response. The main objective of an antenna design for these observations is to achieve a low chromaticity, which translates to having a beam pattern that varies smoothly with 
frequency. Chromatic antenna beams can couple large angular fluctuations in the Galactic foreground into spectral structures that may interfere with the global cosmological 
21-cm signature. The difference between the minimum and maximum gain measured over a given frequency range characterizes beam chromatic effects. Although this study is not 
often conducted during the general antenna characterization process, it is a vital step in global 21-cm line observations, as beam gain changes close to 1.0\% per MHz would 
result in antenna temperature variations of $\lesssim$ 600 mK \cite{10.1093/mnras/stw2696}.\\

\noindent We propose here a new approach for optimizing and designing antennas for global 21-cm line observations. Our approach is based on finding the optimal geometrical 
dimensions of the antenna that yield our desired frequency bandwidth and gain, while ensuring that the antenna chromaticity is kept to a minimum. To achieve this goal, we use 
a commercial electromagnetic simulation software package in conjunction with the Particle Swarm Optimization (PSO) algorithm, using a merit function based on frequency, beam 
chromaticity, and return loss specifications. As a proof of concept, we have tested our method with two types of antennas: a blade dipole antenna \citep{10.1093/mnras/stw2696}
and a \emph{bow-tie} antenna. We show that this optimization process can indeed improve the performance of a given antenna design, as we find that the bow-tie antenna 
outperforms the blade antenna. Furthermore, we validate these results by characterizing scaled models of the antennas in an anechoic chamber. Our best-performing antenna 
model improves the chromaticity by factor between 2 and 3 with respect to the original EDGES-type, meaning that the chromaticity obtained is less than the observed change and can be 
discarded as a possible systematic error.\\

\section{Antenna design and optimization}

\subsection{Antenna requirements}

We have based our design in the parameter requirements summarized in Table \ref{tab:1} and discussed below.
\begin{table}[h]
        \centering
        \caption{Antenna requirements}
        \begin{tabular}{c|c}
        \hline\hline
            Parameter&Value \\\hline\hline
            Frequency band& 40 - 120 MHz \\
            $S_{11}$& $<$ $-8$ dB\\
            3 dB Beamwidth& 30$^\circ<$ HPBW $<$ 60$^\circ$\\ 
            Chromaticity & RMS$_{\text{NBC}}$ $<$ 1\% (per MHz) \\\hline\hline
        \end{tabular}
        \label{tab:1}
    \end{table}

\subsubsection*{Frequency band}
    
    The operating frequency band has to be consistent with the expected redshift/frequency range of the 21-cm cosmological line spectrum. We have, therefore, selected the 50--100~MHz frequency range corresponding to $30 < z < 13$.
    
\subsubsection*{$S_{11}$}

    The antenna needs to have a low reflection coefficient in order to efficiently couple the signal with the rest of the instrument. 
    
    Traditionally, the bandwidth is defined as the frequency range in which the reflection coefficient $S_{11}$ remains below $-10$~dB. This ensures that the antenna has a better level of impedance matching and improves its gain. Since we also intend to 
    optimize over the central frequency and the chromaticity, we have chosen to relax the $S_{11}$ limit to $-8$~dB in order to make it easier for the algorithm to achieve a 
    compromise between the two optimized parameters. We compensate for this limit change by keeping resistive losses low in other instrument subsystems. In the case of a 
    single dipole (such as a blade), the main contribution to resistive losses comes from the length of the cable connecting the antenna panels and the receiver, which can be 
    kept as short as possible. Additionally, the instrument will incorporate an antenna-receiver balun coupling which limits the common mode current and therefore reduces SWR 
    losses.

\subsubsection*{HPBW}

The science requirements on the wavelength range restrict single-antenna experiments to wide-beam, low resolution\footnote{We should note that in this study we define low angular resolution antennas as those with a half-power beamwidth (HPBW) $>$ 30$^\circ$,} antennas such as dipoles. A wide beam antenna can observe a large area of the sky when pointing the radiation pattern towards the zenith ($\theta < 60$ degrees) while trying to avoid radio frequency interference (RFI) and ground emission. This requirement is necessary at all frequencies under study. Any advantage gained from using an antenna array to create a narrower beam is balanced with the higher cost and complexity required for modelling \citep{2022JAI....1150001C}. For these reasons we focus on wide-beam antenna design.

\subsubsection*{Chromaticity}

Antenna chromaticity effects have to be minimized. In order to examine the chromaticity of the antenna beam gain 
\cite{10.1093/mnras/stw2696}, we calculate the derivative of the beam gain with respect to frequency over the $(\nu_i,\theta_j)$ grid \cite{Mahesh_2021} 
at each $\phi=0^\circ$ and $\phi=90^\circ$ to obtain the Beam gain Change (BC) as

\begin{equation}
    \text{BC} = \left(\frac{\Delta G}{\Delta \nu}\right)_{i,j} = \frac{G(\nu_{i+1},\theta_j)-G(\nu_i,\theta_j)}{\nu_{i+1}-\nu_i}\ .
\end{equation}

\noindent In order to better quantify this assessment from our simulations, we use a point-to-point Normalized Beam gain Change (NBC) defined as,

\begin{equation}
    \text{NBC}_{i,j}=\frac{(\frac{\Delta G}{\Delta \nu})_{i,j}}{\overline{G}_{i,j}} *100\%,
\end{equation}

\noindent where $\overline{G}_{i,j} = \frac{1}{2}\left[G(\nu_i,\theta_j)+G(\nu_{i+1},\theta_j)\right]$. We then take the root mean square (RMS) of the NBC grid as a proxy parameter for estimating the overall chromaticity calculated in the whole frequency range and Zenith angle ranging from 0 to 90 degrees to obtain RMS$_{\text{NBC}}$ for each of the $\phi=0^\circ$ and $\phi=90^\circ$ planes. One of our main optimization goals is therefore to minimize the RMS$_{\text{NBC}}$ of the beam map to validate candidate antennas for global 21-cm observations.
  
\subsection{Antennas under study}

    \subsubsection{Blade antenna}
    
        We started from the planar dipole antenna used by EDGES that works in the 100--200 MHz frequency range. This dipole-type blade antenna consists of two 
        rectangular panels fed between the two halves \cite{10.1093/mnras/stw2696} as shown in Figure~\ref{fig:geoblade_v10}. 
        We have scaled its physical dimensions by a factor of 2 to cover the band between 50--100 MHz.
        
\subsubsection{Bow-tie antenna}

\begin{figure}[h]
                \centering
                \begin{subfigure}[b]{0.3\textwidth}
                \centering
                \includegraphics[width=\textwidth]{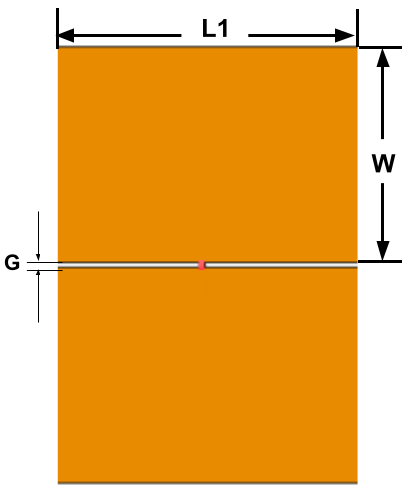}
                \caption{}
                \label{fig:geoblade_v10}
                \end{subfigure}
                \hfill
                \begin{subfigure}[b]{0.36\textwidth}
                \centering
                \includegraphics[width=\textwidth]{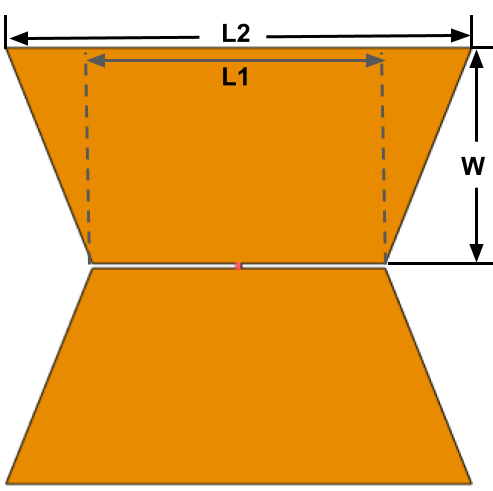}
                \caption{}
                \label{fig:geobowtie_v11}
                \end{subfigure}
                \hfill
            \caption{Schematics of the (a) blade and (b) bow-tie antennas with their 
            geometrical parameters. The antennas have panels of a thickness~$t$ and were studied using a 
            metallic ground plane situated at a distance~$H$.}
            \label{fig:schematic_blade_bowtie_antenna}
        \end{figure}

        The bow-tie antenna is a variation of the blade antenna in which each blade has an oblique angle. This angle is formed by 
        including a new length (L2) in the geometry of the blade antenna (Fig.~\ref{fig:geobowtie_v11}). Defining the antenna structure with an oblique angle is one way to achieve a larger bandwidth \cite{inproceedings}. Moreover, this oblique angle could improve the chromaticity of the antenna.\\  

\subsection{Antenna optimization procedure} 

            We have implemented the PSO method in Python to improve the antenna performance 
            \cite{Jarufe}. The PSO algorithm uses a figure 
            of merit (FoM), $f(\vec{x})$, to be minimized over values of a vector $\vec{x}$ conformed by the geometrical parameters to be varied. The FoM should depend on the desired 
            characteristics required for the antenna. In this case, we used a FoM that allows to optimize $S_{11}$ 
            and the chromaticity. We defined the FoM as 
            
            \begin{equation}
              f(\vec{x})=w_1\times \text{RMS}_{\text{BC}} +w_2\left(\frac{\text{BW}_{S_{11}}-\text{BW}_{ideal}}{f_c}\right)^2, 
              \label{ec:1}
            \end{equation}
            
            \noindent where BW$_{S11}$ is the bandwidth at $-8$~dB of the $S_{11}$ parameter as a function of frequency. BW$_{ideal}$ is the desired bandwidth (80 MHz) and 
            $f_c$ is the central frequency value. $w_1$ and $w_2$ are weighting coefficients tuned to achieve the desired results. The values of $w_1=0.8$ and $w_2=0.2$ were 
            selected to balance the contributions of beam chromaticity and antenna impedance reflections. 
            Finally, RMS$_{BC}$ corresponds to the RMS of the
            Beam gain Change $(\frac{\Delta G}{\Delta \nu})_{i,j}$ without normalization from the beam gain map as a function of frequency and Zenith angle to save computational time. However, the quantity that we use for beam chromaticity analysis is RMS$_{\text{NBC}}$.\\
            
            \noindent The optimization is a repetitive procedure focused on minimizing the merit function. Two methods were 
            used in the optimization process: penalty and trajectory. The penalty method adjusts the objective function to avoid local minima by adding penalizing functions. 
            The trajectory method changes direction at local minima and switches to ascent, exploring the solution space in the opposite direction to find a maximal point. If 
            found, the algorithm switches back to the descent trajectory, potentially leading to a new minimum. The procedure was implemented in Python in conjunction with 
            Ansoft HFSS which simulates the antenna performance. A Python script is used as an interface between the two packages. At each repetition of the optimization 
            procedure, the antenna structures are simulated by HFSS. To calculate the merit function, Python uses the simulation results from HFSS, and the antenna structures 
            are sorted out based on their fitness. Afterward, the best structures with the lowest merit function are drawn, and the PSO is rerun to calculate  the new 
            structures to be simulated in HFSS. The optimization proceeds until either the maximum number of repetitions (20) or the radiation pattern characteristics are close to the desired values. A flow diagram of the optimization program is shown in Fig. \ref{fig:flow_chart}.

    \begin{center}
    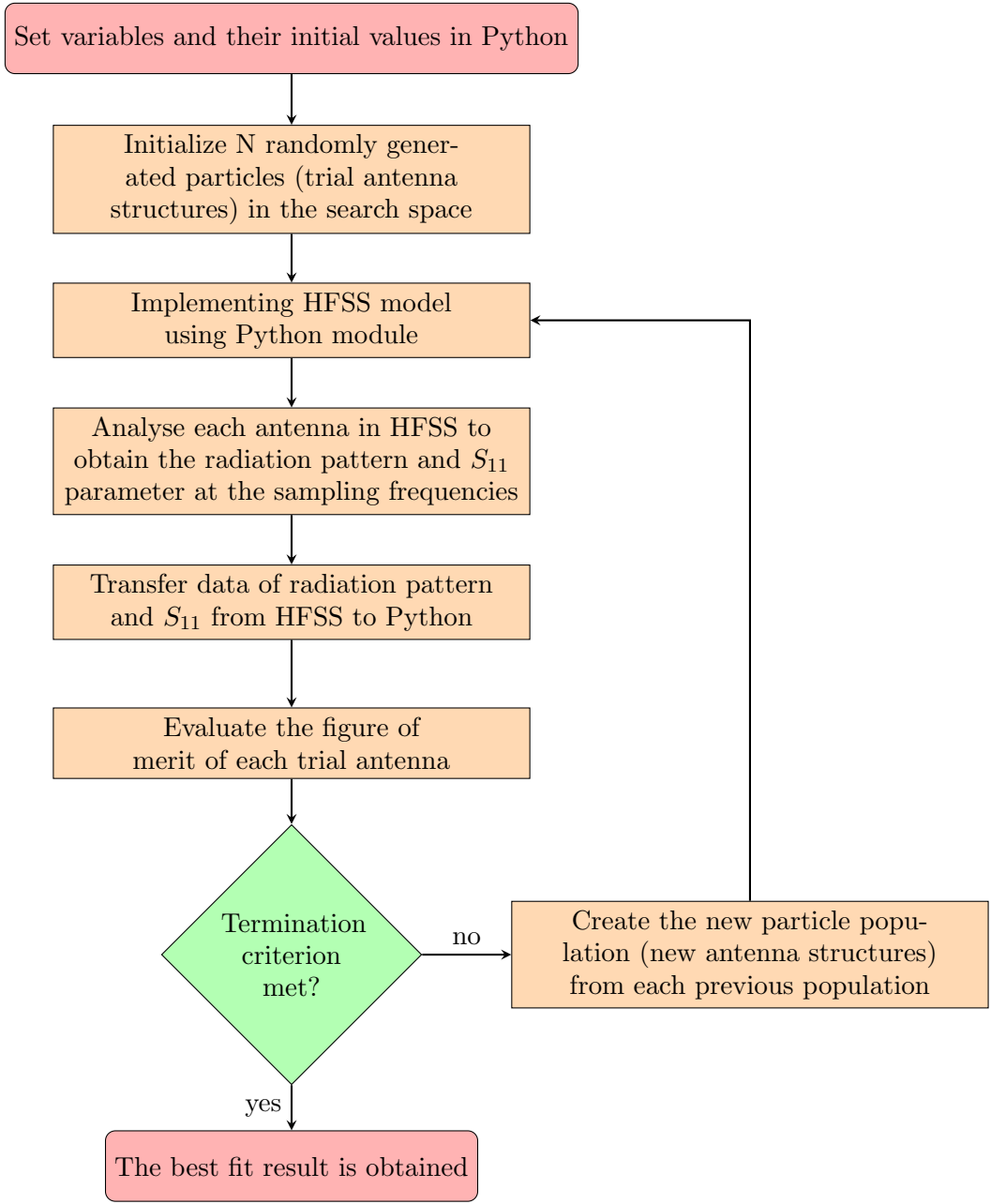
\begin{figure}
        \centering
    \begin{tikzpicture}[node distance=2cm]
    \node (start) [startstop] {Set variables and their initial values in Python};
    \node (pro1) [process, below of=start] {Initialize N randomly generated particles (trial antenna structures) in the search space};
    \node (pro2) [process, below of=pro1] {Implementing HFSS model using Python module};
    \node (pro3) [process, below of=pro2] {Analyse each antenna in HFSS to obtain the radiation pattern and $S_{11}$ parameter at the sampling frequencies};
    \node (pro4) [process, below of=pro3] {Transfer data of radiation pattern and $S_{11}$ from HFSS to Python};
    \node (pro5) [process, below of=pro4] {Evaluate the figure of merit of each trial antenna};
    \node (dec1) [decision, below of=pro5, yshift=-1.0cm] {Termination criterion met?};
    \draw [arrow] (start) -- (pro1);
    \draw [arrow] (pro1) -- (pro2);
    \draw [arrow] (pro2) -- (pro3);
    \draw [arrow] (pro3) -- (pro4);
    \draw [arrow] (pro4) -- (pro5);
    \draw [arrow] (pro5) -- (dec1);
    \node (stop) [startstop, below of=dec1, yshift=-1.0cm] {The best fit result is obtained};
    \node (pro2b) [process, right of=dec1, xshift=4.5cm] {Create the new particle population 
    (new antenna structures) from each previous population};
    \draw [arrow] (pro2b) |- (pro2);
    \draw [arrow] (dec1) -- node[anchor=east] {yes} (stop);
    \draw [arrow] (dec1) -- node[anchor=south] {no} (pro2b);
    \end{tikzpicture}
        \caption{Flowchart of the optimization process.}
        \label{fig:flow_chart}
\end{figure}
\end{center}
            
\section{Results of the antenna optimization}

        All physical dimensions of the antenna, including the thickness of the panels, were varied to achieve a broad bandwidth and a beam as achromatic as possible. The size of the ground plane was the only fixed parameter. The original and optimized dimensions obtained for both blade and bow-tie antennas are summarized in Table~\ref{Table:Blade_2}.\
        
         \begin{table}[h]
                \centering
                \caption{Physical dimensions of the original and optimized blade and bow-tie antennas. All antennas have a 5~m~$\times$~5~m ground plane.}
                 \begin{tabular}{cp{2.4cm}p{2.5cm}p{2.5cm}p{2.6cm}}
                    \hline\hline
                    \centering Physical dimensions  & \multicolumn{4}{c}{Blade\hspace{4.5cm} Bow-tie}\\ & Original (m) &Optimized (m) &Original (m)&Optimized (m)\\\hline\hline
                     W & \makebox[2.2cm][c]{0.98}  &\makebox[2.2cm][c]{1.10}&\makebox[2.2cm][c]{1.00}&\makebox[2.2cm][c]{1.10}\\  
                     L1& \makebox[2.2cm][c]{1.26}&\makebox[2.2cm][c]{1.35}&\makebox[2.2cm][c]{1.22}&\makebox[2.2cm][c]{1.32}\\
                     L2& \makebox[2.2cm][c]{NA} &\makebox[2.2cm][c]{NA}&\makebox[2.2cm][c]{2.10}&\makebox[2.2cm][c]{1.80}\\
                     G & \makebox[2.2cm][c]{0.04} & \makebox[2.2cm][c]{0.02}&\makebox[2.2cm][c]{0.04}&\makebox[2.2cm][c]{0.02}\\
                     t & \makebox[2.2cm][c]{0.006}  &\makebox[2.2cm][c]{0.006} &\makebox[2.2cm][c]{0.006}&\makebox[2.2cm][c]{0.006}\\  
                     H & \makebox[2.2cm][c]{0.75} &\makebox[2.2cm][c]{0.78}&\makebox[2.2cm][c]{0.60}&\makebox[2.2cm]
                     [c]{0.80}\\   
                
                    \hline\hline

              \end{tabular}
              \label{Table:Blade_2}\\
            \end{table}

\subsection{$S_{11}$}
\begin{figure}[h]
                \centering
                \includegraphics[scale=0.4]{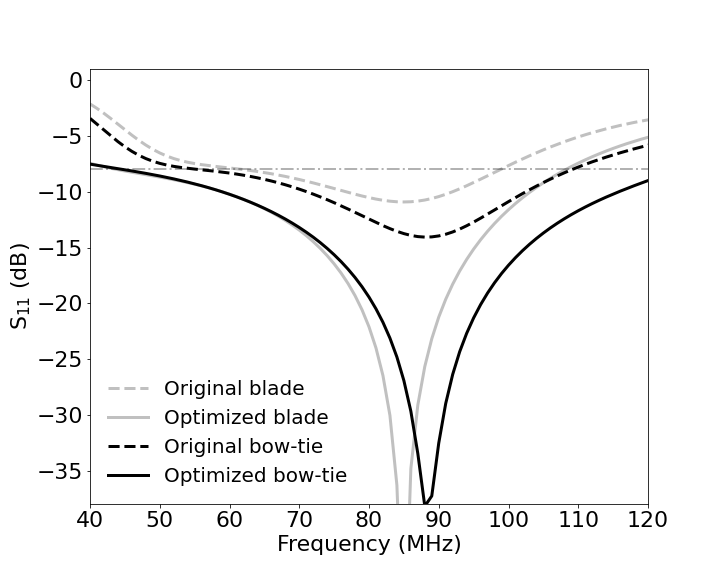}
                \caption{HFSS simulations of $S_{11}$ for the original and optimized antennas.}
                \label{fig:Optimized_ All_S11_v12}
            \end{figure}

        \noindent Fig.~\ref{fig:Optimized_ All_S11_v12} compares the simulated $S_{11}$ parameter of the original and optimized 
        antennas. We can observe that the optimized antennas have a higher bandwidth than the original antennas. This result shows 
        that a desirable $S_{11}$ parameter performance is achieved. For example, the optimized bow-tie antenna shows a bandwidth of 
        78 MHz which, compared to the other antennas, meets the design 
        requirement. In addition, the matching level near 80 MHz is better than in the original antennas.\\
            
            \begin{figure}[h]
                 \centering
                \begin{subfigure}[b]{0.47\textwidth}
                \centering
                \includegraphics[width=\textwidth]{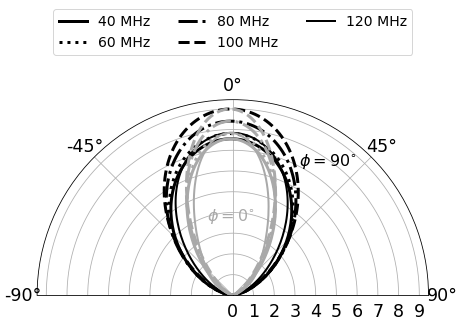}
                \caption{Original blade antenna}
                \label{fig:Gain_phi0_blade_v26}
            \end{subfigure}
            \hfill
            \begin{subfigure}[b]{0.47\textwidth}
            \centering
            \includegraphics[width=\textwidth]{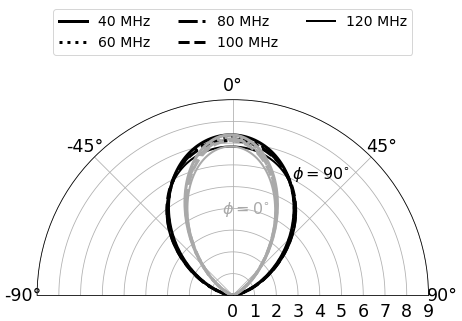}
            \caption{Optimized blade antenna}
            \label{fi:Gain_phi90_blade_v27}
            \end{subfigure}
            \hfill
            \begin{subfigure}[b]{0.47\textwidth}
             \centering
             \includegraphics[width=\textwidth]{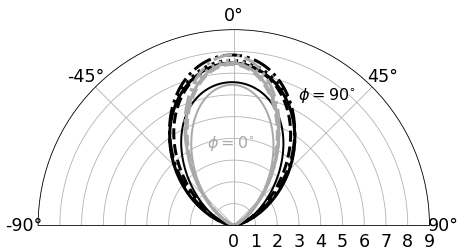}
             \caption{Original bowtie antenna}
             \label{fig:Gain_phi0_optimized_blade_v28}
            \end{subfigure}
            \hfill
            \begin{subfigure}[b]{0.47\textwidth}
            \centering
            \includegraphics[width=\textwidth]{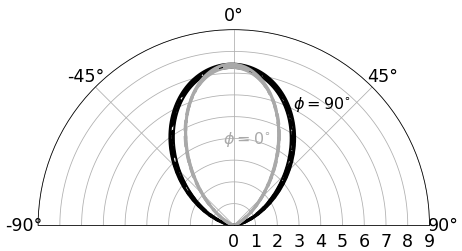}
             \caption{Optimized bowtie antenna}
             \label{fi:Gain_phi90_optimized_blade_v29}
        \end{subfigure}
        \hfill
        \caption{Original (a,c) and optimized (b,d) radiation patterns for the blade and 
        bow-tie antennas. The gain is plotted on a linear 
        scale. Plane $\phi=90^{\circ}$ for cuts along the excitation axis of the blade 
        and bow-tie dipole antennas (E-plane). Plane $\phi=0^{\circ}$ for cuts along the
        perpendicular axis (H-plane).} 
     \label{fig:Gain_both_planes_blade}
    \end{figure}

 \subsection{Beam pattern}
        \noindent Fig. \ref{fig:Gain_both_planes_blade} shows the radiation pattern of the blade and bow-tie antennas. It is notable that the radiation pattern beamwidth for the $\phi$= 0 plane is greater than the $\phi$ = 90 plane. Moreover, the optimized antennas show that the gain spread in the 
        frequency range under study is smaller.\\ 

 \subsection{Beam chromaticity}
    \noindent We also calculated the RMS$_{\text{NBC}}$ of the derivative of the radiation pattern of the antennas before and after optimization. The optimized bow-tie antenna presents lower RMS$_{\text{NBC}}$ values, which means it has a flatter gain than the other geometries. Also, the PSO process decreases the chromaticity by almost 40\% in all cases. Table \ref{Table:Blade_3} summarizes these results. \

    \begin{table}[h]
                \centering
                \caption{Simulation results for bandwidth and chromaticity parameters of the original and optimized blade and bow-tie antennas.}
                    \begin{tabular}{cp{2.4cm}p{2.5cm}p{2.5cm}p{2.6cm}}\hline\hline
                    \centering Parameter  & \multicolumn{4}{c}{Blade\hspace{4.5cm} Bow-tie}\\ & \hspace{0.4cm}Original &\hspace{0.3cm}Optimized &\hspace{0.3cm}Original &\hspace{0.3cm}Optimized 
                    \\\hline\hline                    
                    BW at $- 8 \text{dB}$ (MHz)& \makebox[2.2cm][c]{34.12}&\makebox[2.2cm][c]{64.26}&\makebox[2.2cm][c]{41.27 }&\makebox[2.2cm][c]{76.81}\\
                    RMS$_{\text{NBC}}$ (per MHz)($\phi=0$)& \makebox[2.2cm][c]{0.81 \%}&\makebox[2.2cm][c]{0.42 \%}&\makebox[2.2cm][c]{0.77 \%}&\makebox[2.2cm][c]{0.41 \%}\\
                    RMS$_{\text{NBC}}$ (per MHz)($\phi=90$)& \makebox[2.2cm][c]{1.05 \%}&\makebox[2.2cm][c]{0.63 \%}&\makebox[2.2cm][c]{0.98 \%}&\makebox[2.2cm][c]{0.58 \%}\\\hline\hline
              \end{tabular}
              \label{Table:Blade_3}\\
            \end{table}
    
\section{Validation with scaled antennas}
In order to verify the optimization process and simulations, three antennas were scaled to the 0.8--2.4~GHz range and built: the original and optimized blade antenna, and the optimized bow-tie antenna. However, the scaled antennas were characterized only in the frequency range of 1-2 GHz due to technical reasons.

       \subsection{Scaling and construction} 
               
        \noindent The physical dimensions of $W$, $L1$, $L2$, and $H$ antennas were scaled by a factor of 1/20. 
        However, the thickness of the panels was reduced by only a factor of 2 and $G$ was fixed at 3 mm due to manufacturing 
        constraints. Since the scaling is not 1:1, a direct comparison with figures \ref{fig:Optimized_ All_S11_v12} and 
        \ref{fig:Gain_both_planes_blade} is not possible. Therefore, the antennas were re-simulated with these new values without performing any further optimization.\\
       
\begin{figure}
           \centering
           \includegraphics[scale=0.15]{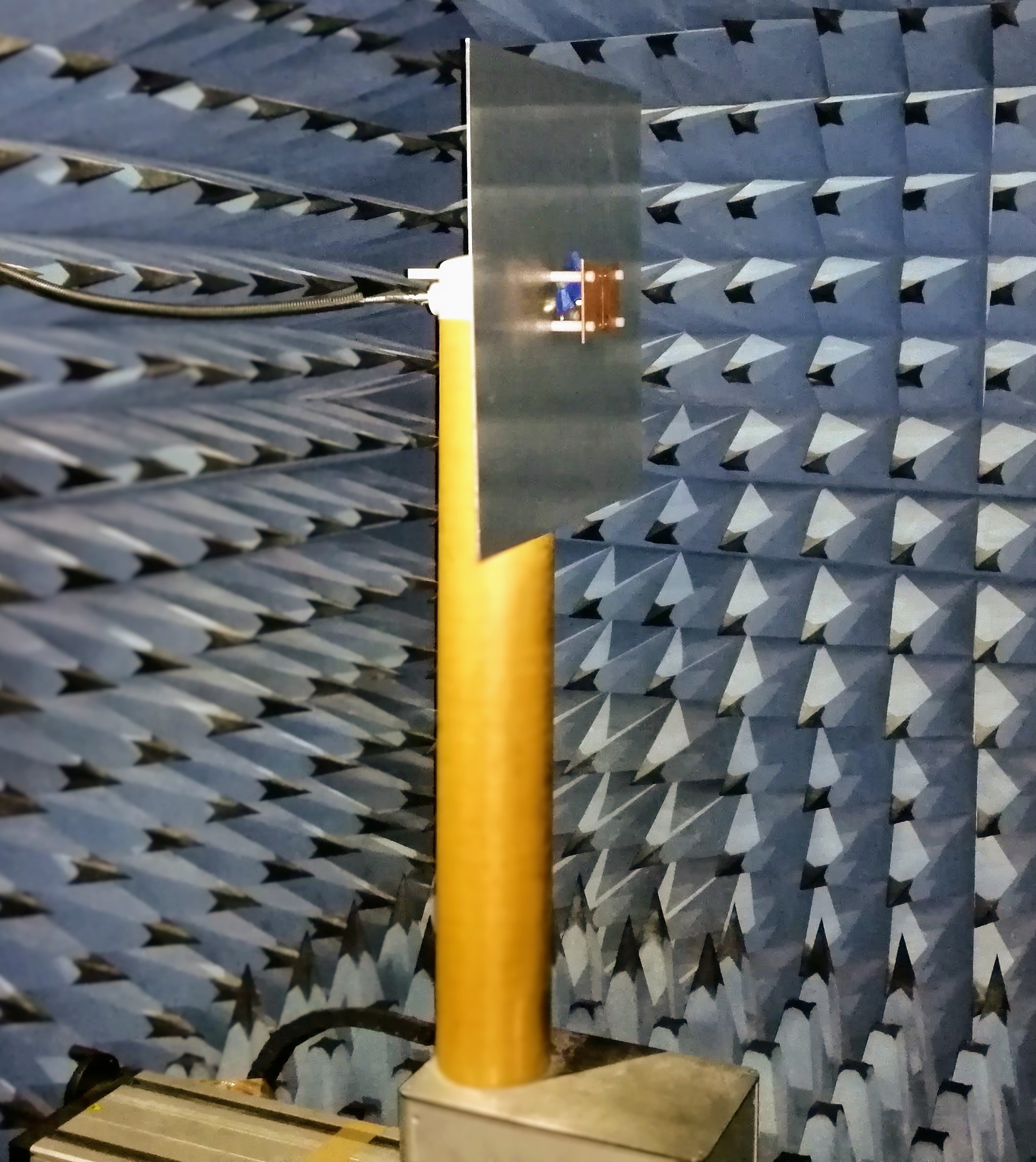}
           \caption{Constructed prototype antenna inside of the anechoic chamber.}
           \label{fig:prototype_scaled}
        \end{figure}

        \noindent A finite metallic ground plane made of aluminum was included. The size of the ground plane was 50 cm $\times$ 50 cm. The antenna panels
        were fabricated from 3-mm thick copper. Since they must be placed at a given height with respect to the ground plane, nylon brackets and 
        bolts were made to hold them and keep them at the proper distance  (see Fig. \ref{fig:prototype_scaled}). The feed 
        was supplied using a female SMA connector connected directly to the panels.\\ 
        
        \subsection{Experimental measurements}
           
           We measured the $S_{11}$ parameter and radiation pattern for the three antennas. The $S_{11}$ of the antennas were measured in the 1--2 GHz range using a Vector Network Analyzer (VNA) Keysight E5061B. Absorbers were used to isolate any external noise source to the antenna.\\ 
           
           \noindent For the antenna radiation pattern measurements, we used an anechoic chamber located at Universidad T\'ecnica Federico Santa 
           Mar\'ia in Valparaiso, Chile. The dimensions of the chamber are 2.5-m high, 3-m wide and 7-m long. The anechoic chamber has a VNA 
           Agilent 8720ES that covers a frequency range between 800~MHz--20~GHz. This setup allowed us to measure the radiation pattern for the H and E planes in the 1--2~GHz range in steps of 50 MHz and 1 degree in elevation angle.\\
                
 \section{Experimental results and discussion}

We present the results obtained from the $S_{11}$ and beam pattern measurements. The beam chromaticity is derived from the beam pattern measurements.

    \subsection{$S_{11}$}    

\begin{figure}[h]
                    \centering
                    \includegraphics[scale=0.49]{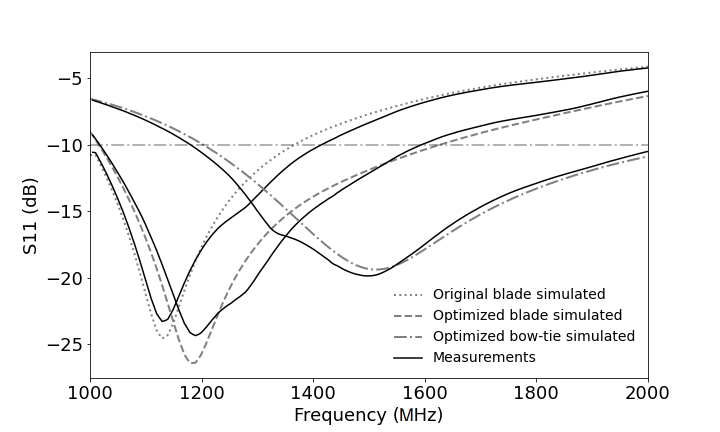}
                    \caption{Comparison between measurements and simulations of the $S_{11}$ parameter for scaled blade and bow-tie antennas.}
                    \label{fig:S11_simulated_scaled}
            \end{figure}

        Fig. \ref{fig:S11_simulated_scaled} shows a good agreement between the results of the simulated and measured $S_{11}$, especially at high frequencies. We found that the optimized blade and bow-tie antennas show a bandwidth, measured at $-$ 10 dB, 
        greater than the original antenna, with the bow-tie antenna having a bandwidth approximately 2 and 1.5 times greater than the original and optimized blade antenna. 
        We attribute the slight differences between simulations and measurements to the antenna feed length (G) and the non-parallelism between panels causing G to be non-uniform. 
        When implementing the antennas, it was not possible to ensure that the feed length was exactly 1 mm and that the panels remained parallel. However, these deviations are small enough to continue validating the design process.
 
    \subsection{Beam pattern}
    
\begin{figure}[h]
        \centering
        \includegraphics[scale=0.27]{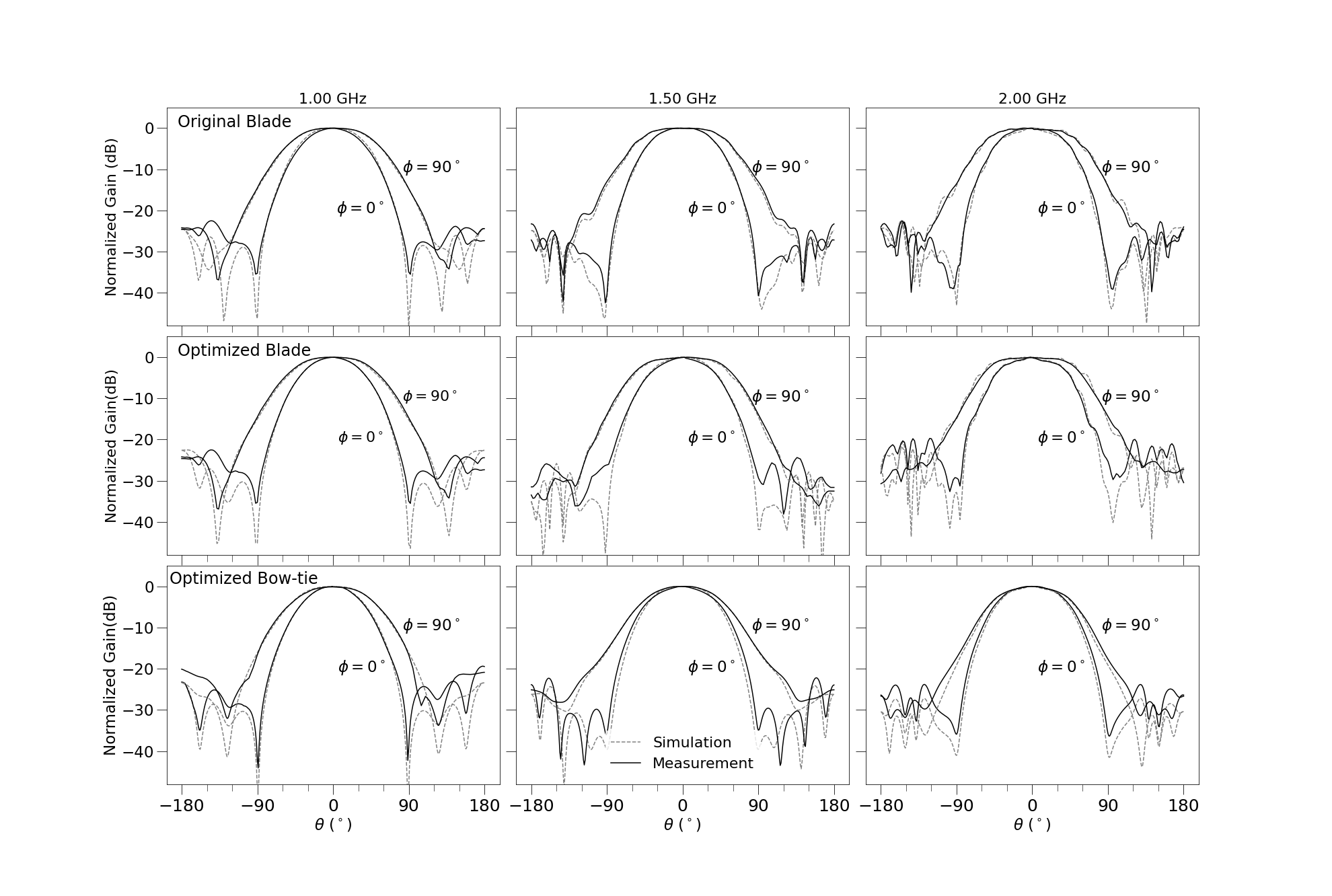}
        \caption{Simulation and measured radiation patterns of the three scaled prototype antennas.}
        \label{fig:blabe_0}    
    \end{figure}

\begin{figure}[h]
        \centering
        \includegraphics[scale=0.62]{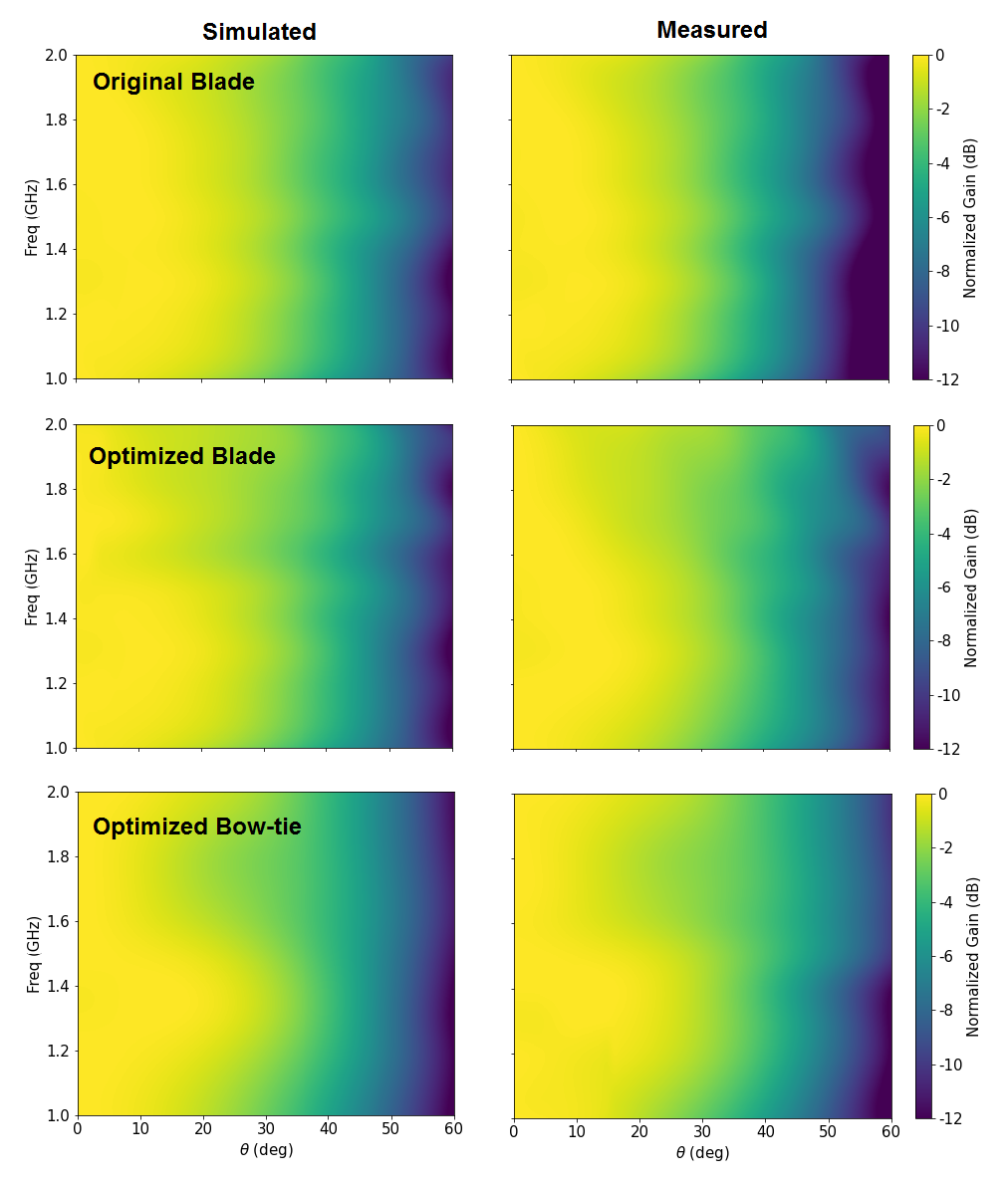}
        \caption{Simulation and measured radiation maps as a function of frequency (y-axis) and Zenith angle $\theta$ (x-axis) for the three scaled prototype antennas. Plane $\phi=0^\circ$.}
        \label{fig:beam_gain_0}    
    \end{figure}

\begin{figure}[h]
        \centering
        \includegraphics[scale=0.62]{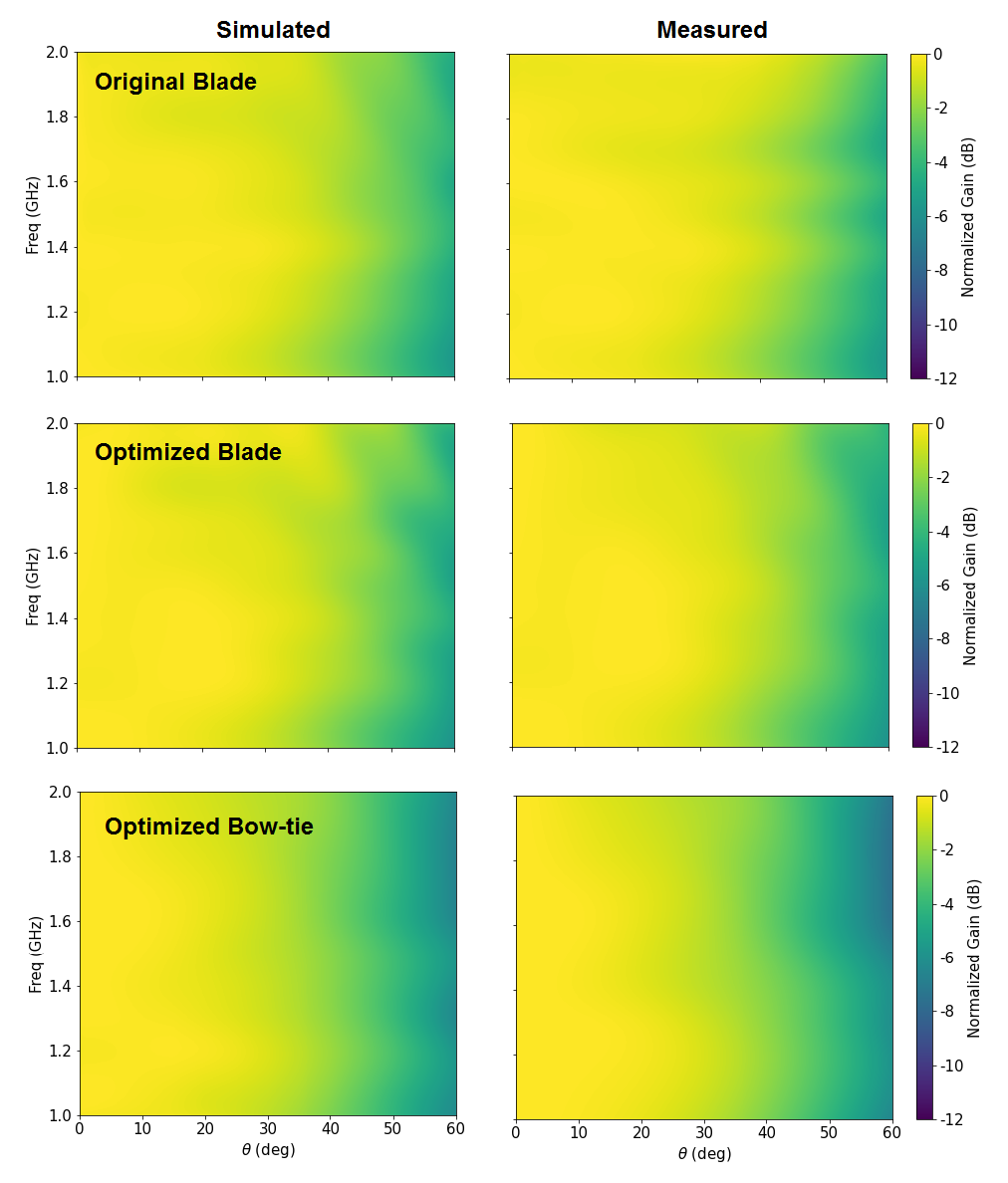}
        \caption{Simulation and measured radiation maps as a function of frequency (y-axis) and Zenith angle $\theta$ (x-axis) for the three scaled prototype antennas. Plane $\phi=90^\circ$.}
        \label{fig:beam_gain_90}    
    \end{figure}

        We performed antenna gain measurements to determine the chromaticity levels for each one of the antennas. The comparison between simulations 
        and measurements for the normalized gain are shown in Fig. \ref{fig:blabe_0}. These figures show good agreement between simulation and 
        measurements, from 0 to approximately $- 20$ dB. The shape of the antenna beam is an ellipse projected 
        on the sky, it has an HPBW of approximately $110 \pm 2$~degrees in the H-plane and $70^\circ \pm 2^\circ$ in the E-plane over the entire studied frequency range. At this beam width, we find that the differences, on average, are less than 0.4 dB. We also see that the agreement remains in 
        the azimuth range from $- 90^\circ$ to $90^\circ$. We can also see that the antennas have their maximum gain for angles close to $0^\circ$. These 
        results indicate that our antennas are pointing at the zenith as expected. For Zenith angles greater than $90^\circ$ and less than $- 90^\circ$, we find larger differences that are attributed, to the measurement capability of the anechoic chamber and to ground plane effects 
        on the radiation pattern, since the antenna is behind the ground plane during measurements. Finally, 3D plots of the gain pattern at various frequencies from simulations and measurements are shown in Figures \ref{fig:beam_gain_0} and \ref{fig:beam_gain_90}. In these cases, we can see a balanced distribution of gain across all frequencies and a smooth performance of the gain, which makes it hard to detect chromatic problems. However, even though it appears smooth in the figures, the antenna may still have chromatic issues at the levels required for this type of experiment. Therefore, it is crucial to carry out a thorough study of the antenna beam behavior.

\subsection{Beam chromaticity}

\begin{figure}[h]
    \centering
    \includegraphics[scale=0.62]{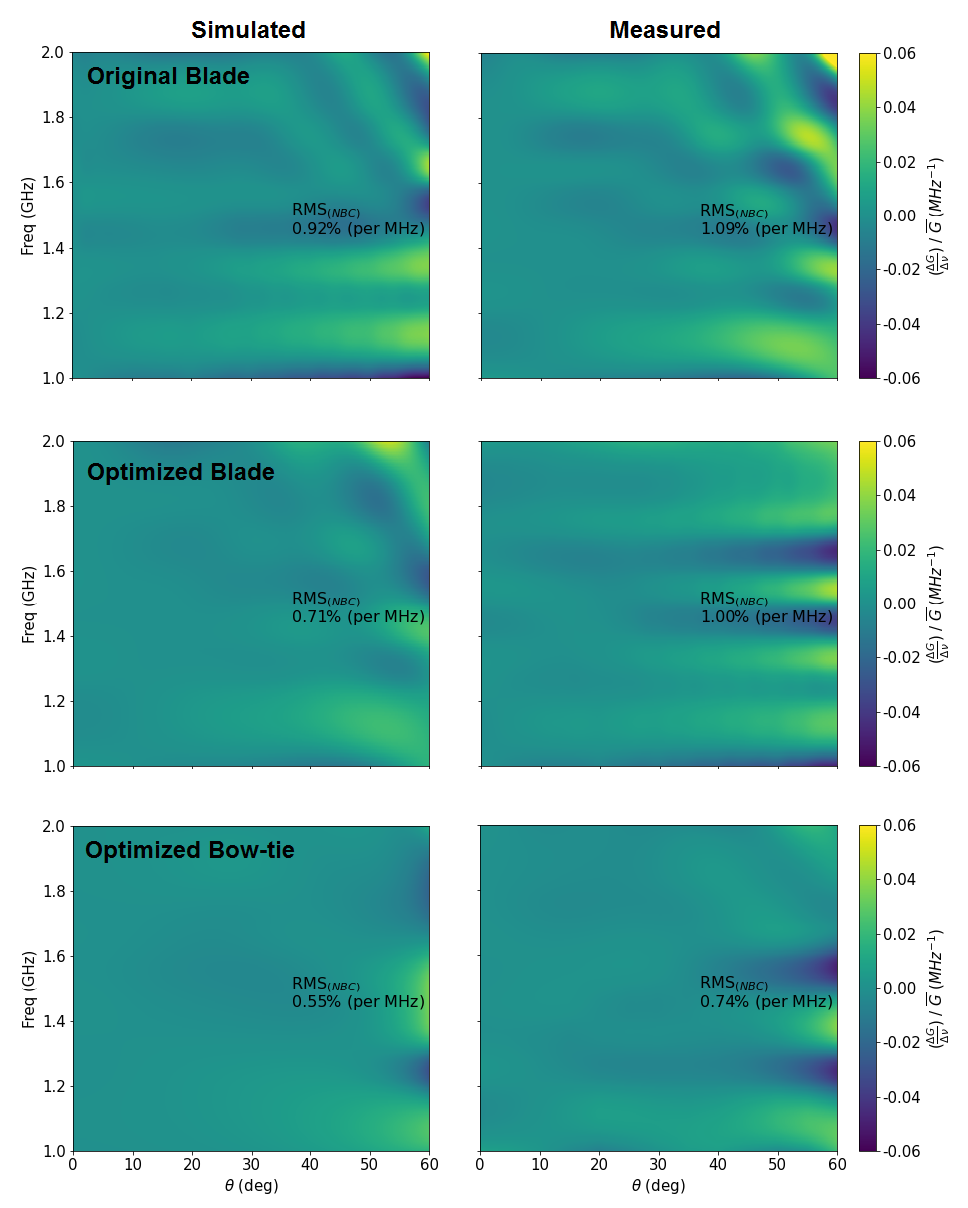}
    \caption{Color plot for the simulated and measured gain derivative as a function of frequency (y-axis) and Zenith angle $\theta$ (x-axis) for 
    the three antennas. The RMS$_{\text{NBC}}$ is calculated for entire band in the plane $\phi=0^\circ$}
    \label{fig:Change_gain_measured_v2_phi_0}    
    \end{figure}

\begin{figure}[h]
    \centering
    \includegraphics[scale=0.62]{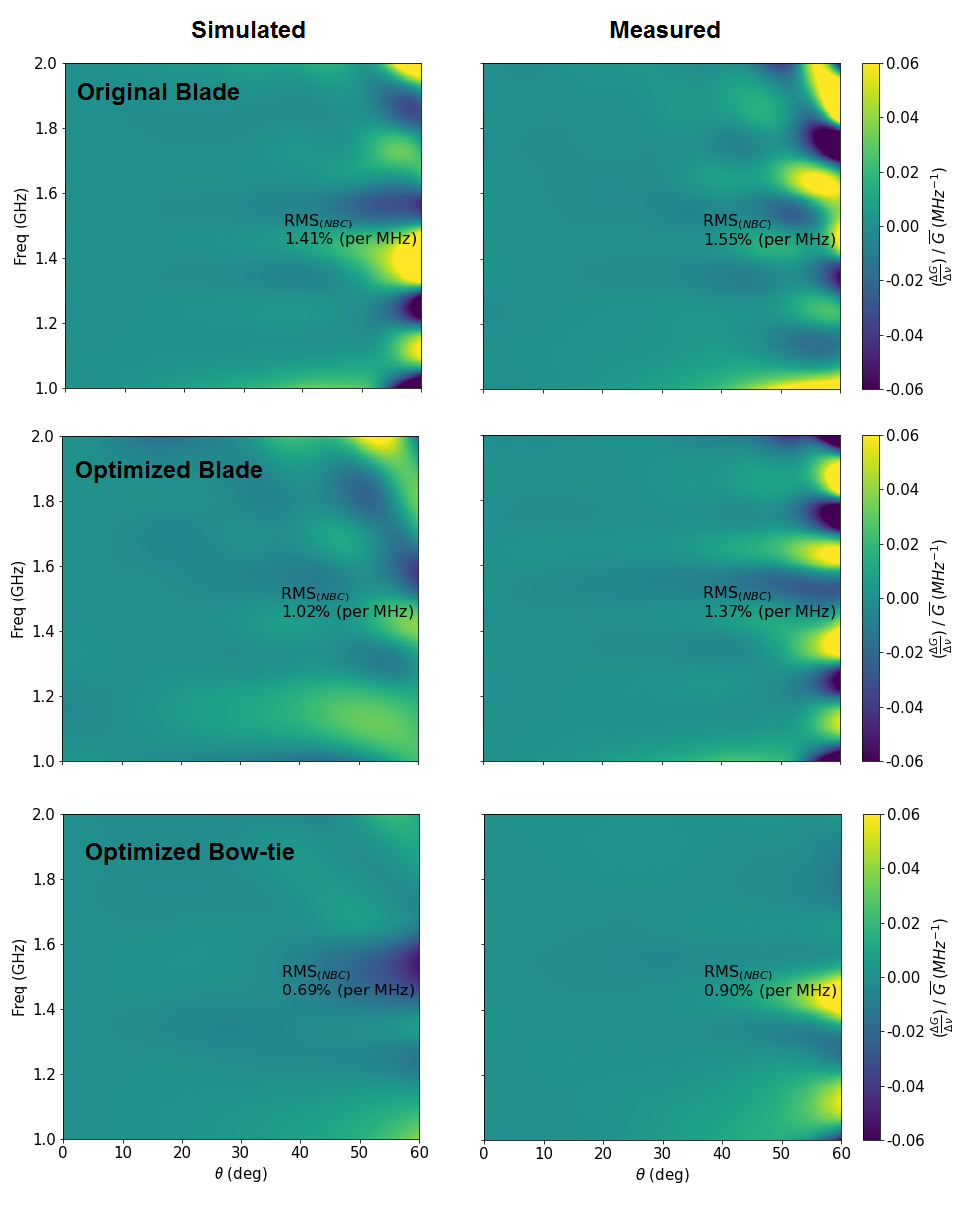}
    \caption{Color plot for the simulated and measured gain derivative as a function of frequency (y-axis) and Zenith angle $\theta$ (x-axis) for 
    the three antennas. The RMS$_{\text{NBC}}$ is calculated for entire band in the plane $\phi=90^\circ$}
    \label{fig:Change_gain_measured_v2_phi_90}    
    \end{figure}

Comparisons between the simulations and measurements regarding gain chromaticity are shown in Figures~\ref{fig:Change_gain_measured_v2_phi_0} 
and~\ref{fig:Change_gain_measured_v2_phi_90}. All antennas show smooth beam gain changes near the zenith for both planes. The original and optimized blade antennas are 
similar in amplitude and chromatic characteristics (structures appearing at elevation angles greater than 20$^\circ$). However, the original blade 
antenna shows more significant changes in magnitude and additional structure at the zenith and elsewhere than the other two antennas. The bow-tie 
antenna shows a much flatter and almost structure-less behavior towards the zenith. The bow-tie antenna has an RMS$_{\text{NBC}}$ close to 0.70 \% (per MHz) indicating 
that this antenna has a better achromatic behavior than the blade antenna. Table \ref{Table:RMS_simulated_vs_measured} summarizes the results of the comparison between 
simulation and measurement of the beam chromaticity parameter for the three antennas. These results show that the performance of the bow-tie antenna is 
improved with respect to the blade antenna. This is in good agreement with the results shown in \cite{2022JAI....1150001C} and \cite{Chinos}.\\ 

      \begin{table}[h]
                \centering
                \caption{Simulated and measured beam chromaticity parameter for the three antennas.}
                \begin{tabular}{cp{1.7cm}cp{1.7cm}cp{1.8cm}cp{1.7cm}}
                    \hline\hline
                    Antenna parameters  & \multicolumn{6}{c}{original Blade\hspace{2.0cm} 
                    Optimized Blade\hspace{2.0cm} Optimized Bow-tie}\\ & \hspace{0.5cm}$\phi=0^\circ$ &$\phi=90^\circ$ 
                    &\hspace{0.5cm}$\phi=0^\circ$&$\phi=90^\circ$&$\hspace{0.5cm}\phi=0^\circ$&$\phi=90^\circ$\\\hline\hline
        RMS$_{\text{NBC}}$ (simulated)&\makebox[2.0cm][c]{0.92 \%}&\makebox[2.2cm][c]{1.41 \%}&\makebox[2.0cm][c]{0.71 \%}&\makebox[2.2cm][c]{1.02 \%}&\makebox[2.0cm][c]{0.55 \%}&\makebox[2.2cm][c]{0.69 \%}\\
        RMS$_{\text{NBC}}$ (measured)& \makebox[2.0cm][c]{1.09 \%}&\makebox[2.2cm][c]{1.55 \%}&\makebox[2.0cm][c]{1.00 \%}&\makebox[2.2cm][c]{1.37 \%}&\makebox[2.0cm][c]{0.74 \%}&\makebox[2.2cm][c]{0.90 \%}\\   
                    \hline\hline
               \end{tabular}
               \label{Table:RMS_simulated_vs_measured}  
            \end{table}

 \section{Conclusions}
 
 We have demonstrated that the PSO method has improved the overall performance of blade and bow-tie antenna models respect to bandwidth 
 and chromaticity for cosmological 21-cm line observations. We found that the chromaticity effects observed in the blade antenna were 
 reduced by optimizing its dimensions via the PSO method. We also demonstrated that the optimized bow-tie antenna presents better 
 bandwidth and lower chromaticity with respect to the blade antenna. We validated the simulation results by directly characterizing 
 scaled-down versions of the antenna models, showing a very good match between our simulations and the antenna response measurements. Overall, the results of our work demonstrate that the presented optimization process 
 is effective and offers improved performance in terms of bandwidth and chromaticity for both blade and bow-tie antennas in observations of the 21 cm line in cosmology.
 \\

\section*{Acknowledgments}

We acknowledge support from ANID Chile Fondo 2018 QUIMAL/180003, Fondo 2020 ALMA/ASTRO20-0075, Fondo 2021 QUIMAL/ASTRO21-0053, UCSC 
BIP-106 and Proyecto "USC20102 Internacionalización Transversal en la UCSC: enfrentando los nuevos desafíos". OR thanks Universidad ECCI 
for financial support via convocatoria interna 06-2020 Vicerrectoría de Investigación. We would like to thank Universidad 
T\'ecnica Federico Santa Mar\'ia for the use of the anechoic chamber. And, we also like to thank Raúl Monsalve, H. Cynthia Chiang, Jonathan Sievers, Matheus Pessoa, and Ian Hendricksen.

\bibliographystyle{ws-jai}

\bibliography{sample}

\end{document}